\documentclass[conference]{IEEEtran}
\usepackage{graphics}
\usepackage{pbox}

\ifCLASSINFOpdf
   \usepackage[pdftex]{graphicx}
\else
   \usepackage[dvips]{graphicx}
\fi
\begin{document}
%
\title{A Taxonomy of Malicious Traffic for Intrusion Detection Systems}



\author{\IEEEauthorblockN{Hanan Hindy\IEEEauthorrefmark{1},
Elike Hodo\IEEEauthorrefmark{3}
Ethan Bayne\IEEEauthorrefmark{1},\\
Amar Seeam\IEEEauthorrefmark{2}, 
Robert Atkinson\IEEEauthorrefmark{3} and
Xavier Bellekens\IEEEauthorrefmark{1}}

\IEEEauthorblockA{\IEEEauthorrefmark{1}Division of Cyber Security, Abertay University, Dundee Scotland}
\IEEEauthorblockA{\IEEEauthorrefmark{2}Department of Computer Science, Middlesex University (Mauritius)}
\IEEEauthorblockA{\IEEEauthorrefmark{3}EEE Department, University of Strathclyde, Glasgow, Scotland}}


\maketitle

\begin{abstract}
With the increasing number of network threats it is essential to have a knowledge of existing and new network threats in order to design better intrusion detection systems. In this paper we propose a taxonomy for classifying network attacks in a consistent way, allowing security researchers to focus their efforts on creating accurate intrusion detection systems and targeted datasets. 
\end{abstract}

\IEEEpeerreviewmaketitle

\section{Introduction}
Cyber-security is defined as the discipline concerned with protecting networks, computer devices, programs and data from different forms of attacks. Cyber-attacks could cause data loss, allow attackers to access confidential information or affect system or service availability. Research in this domain focuses on detecting the attacks and on preventing and predicting them. Although research in this field started in the early fifties, it is an evolving domain~\cite{tikk2015evolution}. 
\\\\
The importance of research in this domain grow with the increasing prevalence of Internet of Things (IoT)  systems to aggregate, transfer and send data to and from sensors. It is predicted by CISCO that there will be 50 billion devices connected to the Internet by 2020 \cite{evans2011internet}. 

Attacks are becoming more complex, targeting a wide range of devices (industrial, personal, etc)~\cite{bellekens2015cyber}. Some attacks focus on obtaining information without causing any damage to a system, whilst some attacks cause damage either by manipulating information or masquerading the attack to access privileges maintain an access. Moreover, users are becoming more aware of attacks and as a result systems are now designed to be  more secure~\cite{krister2009automated}. Therefore, the current iteration of strategic analysis tools and methods will no more fit the need for threats detection and prevention.
The taxonomy of malicious traffic presented in this paper can help researchers and software engineers to design up-to-date detection tools and to find ways to prevent and predict these attacks. 
\\\\
The remainder of the paper is organised as follows, Section~\ref{problem} highlights the need for new taxonomy focused on network threats. Section~\ref{RW} provides an insight on related work, while in Section~\ref{Taxonomy} the taxonomy is described and analysed, finally, the paper summarises with the conclusion in Section~\ref{conclusion}.

\section{Problem Statement}
\label{problem}
Current rule based intrusion detection system only consider a subset of known attacks to defend large enterprise networks. Rule based intrusion detection systems rely highly on prior detection of attacks and regularly updated rules~\cite{bellekens2014glop}. While these systems provide a first step to security, they do not enable the detection of unknown attacks. Machine learning intrusion detection systems on the other hand allow the detection of unknown threats, however these system rely heavily on existing training datasets, that may be outdated and leave out a number of recent threats~\cite{hodo2017machine}. This problem  underpins the need for a taxonomy of attacks allowing researchers and engineers to build better datasets~\cite{hodo2017shallow}.
\section{Related Work}
\label{RW}
The increasing number of threats has led to advances in cyber-security. These systems however have numerous drawbacks leading to systems being compromised. To this end, researchers have published a number of taxonomies with the aim to increase the overall efficacy of threat detection systems. 
\\\\
Zhu~\textit{et al.}~\cite{zhu2011taxonomy} provide a taxonomy detailing the different flaws and attacks against industrial SCADA systems. The attacks are classified using the TCP/IP stack. Chakrabarti~\textit{et al.} highlight the threats the Internet infrastructure is facing~\cite{chakrabarti2002internet} the attacks are illustrated through different scenarios. 
Hoque~\textit{et al.}~\cite{hoque2014network} provides a taxonomy of tools and systems against network attacks, detailing the numerous tools used to identify network flaws. The tools are described and then classified with their pros and cons  described.
\\\\
While these taxonomies provide information on network attacks, they focus on specific systems and tools. The taxonomy presented in this paper focuses primarily on threats that can be detected via an intrusion detection systems and is aimed at researchers building datasets to ensure that relevant attacks are included, hence, increasing the efficiency of future intrusion detection systems that incorporate  machine learning.

\begin{figure*}[t]
\centering
\includegraphics[width=\textwidth]{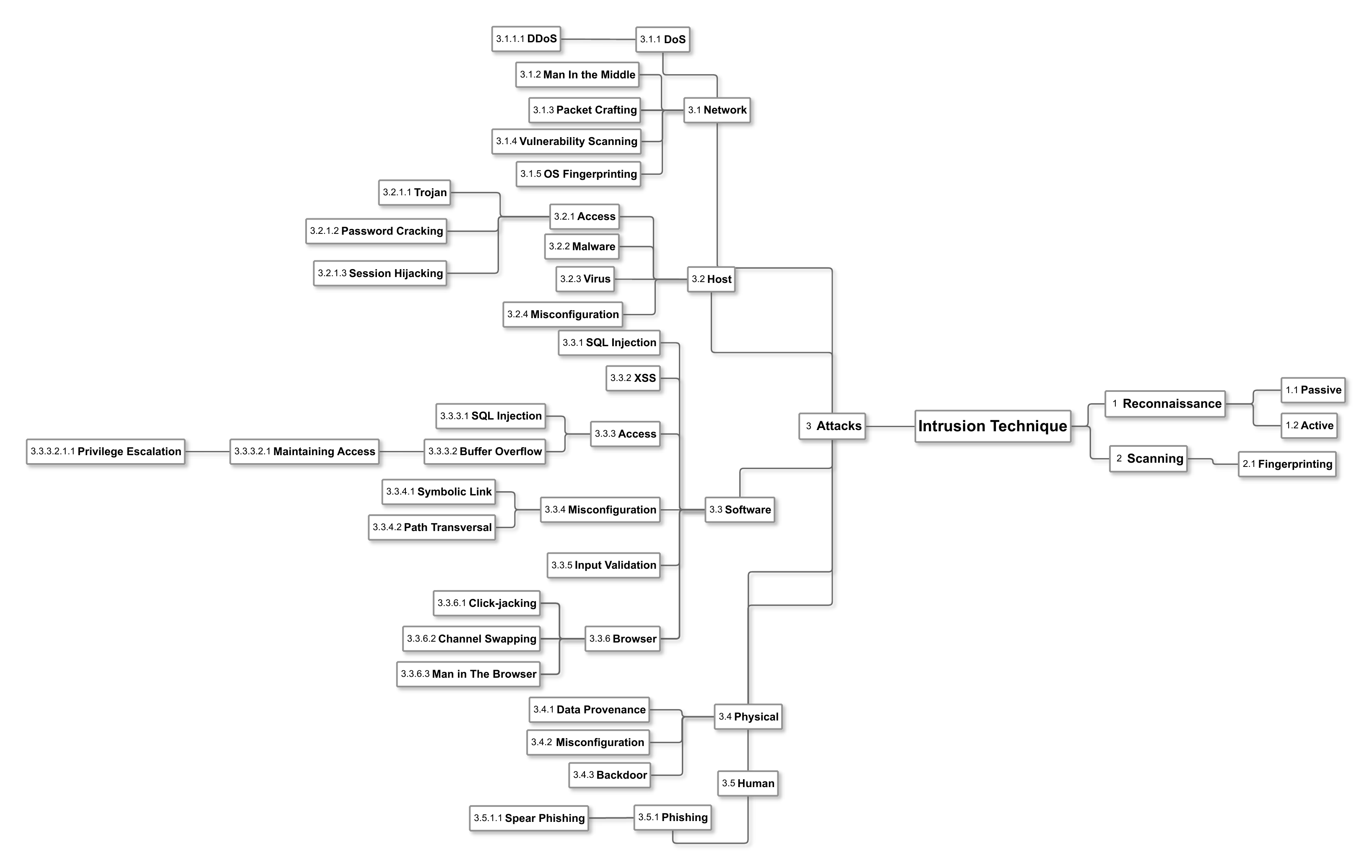}
\caption{Malicious Traffic Taxonomy}
\label{Fig:TaxonomyIntrusion}
\end{figure*} 

\section{Taxonomy of Malicious Traffic}
\label{Taxonomy}
The aim of this manuscript is to help researchers and engineers to develop techniques to detect current and new malicious traffic occurring on the network.The taxonomy presented in this work is composed of three control stages (CS) representing attack steps taken by malicious users. Each stage is then further composed of sub-stages that can be either executed in parallel or subsequently during an attack. 
An example of multiple sub-steps being subsequently being executed is a malware taking advantage of a physical host executing a buffer overflow in order to provide access to a malicious user. 
\\\\
The taxonomy provided is designed to provide the end-user with singular sub-components that can be added together to represent a more complex attack. \newline
\textbf{Control Stage 1 : Reconnaissance}
The reconnaissance stage is often the initial and the most important stage~\cite{wrightson2014advanced}. It is used by malicious users to gather data on the target. The data gathered by the reconnaissance inform both CS2 and CS3. This stage enables the malicious user to learn critical information about the target, through various passive and active reconnaissance and obtain specific data such as 
\begin{itemize}
\item IP Address Range
\item DNS Records
\item Mail Servers 
\end{itemize}
\textbf{Control Stage 2: Scanning}
Malicious users can gather critical information about the network target by mapping the system and the network. Important network components such as firewalls, routers, etc, can be discovered. This task is often realised by using a port scanner. The port scanner is designed to try and target opened ports. Scanning, is an active step, to identify network services running on a host; it also allows the malicious user to test the network and the firewall for the different security policies in place. This stage informs CS3. 

\begin{table}[b]
\caption{Port Scanning Responses}
\label{Control Stage 2: Port Scanning Responses}
\begin{tabular}{|l|l|}
\hline
\textbf{Category} & \textbf{Description} \\ \hline
Opened & \pbox{10cm}{The target responded on a that the port,\\ implying that a particular service is listening} \\ \hline
Closed & \pbox{10cm}{The target responded that all connections\\ are denied on that port} \\ \hline
Filtered / Blocked & \pbox{10cm}{the target did not reply} \\ \hline
\end{tabular}%
\end{table}

\textbf{Control Stage 3 : Attacks}
Whilst CS1 and CS2 can be operated with relative stealth on the network, the attack is essentially critical for the malicious user as both CS1 and CS2 inform the choice of target (e.g. which server represents the best target), which of exploit should be used to obtain the best result, etc. Figure~\ref{Fig:TaxonomyIntrusion} provides a taxonomy of malicious attacks that can be detected by analysing network traffic using an intrusion detection system or deep packet inspection. The attacks are classified in five different subsections described hereafter;
\subsection{Network}
Network nodes are vulnerable to a large range of attacks. Figure~\ref{Fig:TaxonomyIntrusion} (3.1) depicts common attacks on a network ecosystem. DoS and DDoS attacks (3.1.1) and (3.1.1.1) can be characterised by the large number of packets or requests received by target with an intent to render it's main function unusable (e.g. Sending a high number of HTTP request on a web server in an attempt to overwhelm the server and discard legitimate connections)~\cite{mirkovic2004taxonomy}. The Man in The Middle attack (3.1.2) (e.g. ARP flooding attack). This attack attempts to map the wrong MAC address with an IP address, allowing the malicious user to redirect incoming and outgoing traffic and snoop on the network conversation~\cite{callegati2009man}. The packet crafting attacks (3.1.3) highlights the possibility for an attacker to replay or craft a new packet to bypass a firewall, test the TCP/IP stack of network components, or replay a packet to gain access to a network component~\cite{papp2015embedded}.  The vulnerability Scanning (3.1.4) is composed of all the methods used by malicious attackers to scan a network for vulnerabilities with known scanners such as Nessus, NMAP, OpenVAS, MBSA etc. Vulnerability scanners can allow malicious user to detect vulnerabilities and how to exploit them~\cite{holm2012performance}. 
\subsection{Host}
As shown in Figure~\ref{Fig:TaxonomyIntrusion} (3.2), Host based threats encompass infection of a host or its access through malicious intent. Gaining Access (3.2.1) describes multiple methods to gain illegal access using trojans (3.2.1.1) or backdoors for remote access~\cite{boraten2018mitigation}, password cracking (3.2.1.2) using tools such as john the ripper to gain access to the target~\cite{mahey2018graphical}, or session hijacking (3.2.1.3) to access the session of a legitimate user using a replay attack~\cite{shahriar2018fuzzy}. Host based (3.2) attacks also describe malware attacks (3.2.2) describing adware, sypware and worms, as well as viruses (3.2.3) which is in itself a malware, however has the ability to infect other hosts on the network during execution. 

\subsection{Software}
Figure~\ref{Fig:TaxonomyIntrusion} provides an overview of software based attacks (3.3) that can be detected in network traffic. (3.3.3.1) SQL Injection are commonly used to maliciously acquire data from a database~\cite{halfond2006classification}, (3.3.3.2) Buffer Overflow can be used to access restricted data in memory, they can also be used to gain root access on a computer, or for privilege escalation (3.3.3.2.1) in order to maintain access on the target host~\cite{pasupulati2004buttercup}. Software can also be subject to misconfiguration such as the symlink attack (3.3.4.1) on web servers or path transversal attacks (3.3.4.2) that are used to access folders and files outside the web root folder. Software can also be vulnerable to input validation (3.3.5) attacks~\cite{simmons2006preventing}. This attack requires the malicious user to test different input fields to obtain error messages and gain access either to a system, or to execute code to obtain data.

\begin{figure}[t]
\centering
\includegraphics[width=2in]{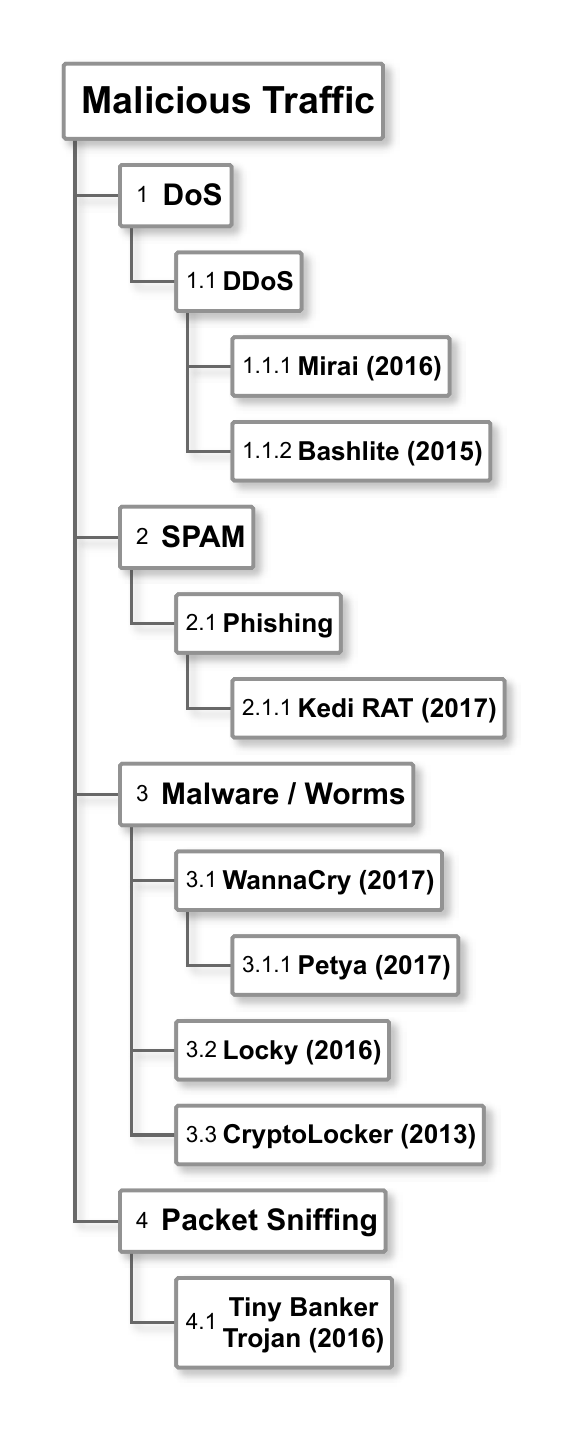}
\caption{Malicious Traffic Including Exploit and Worm Names}
\label{Fig:Examples}
\end{figure} 

\subsection{Physical}
Network nodes are vulnerable to large range of physical attacks, a number of these attacks are highlighted in Figure~\ref{Fig:TaxonomyIntrusion} (3.4). Physical attacks include physical back-doors added in foundries that can be enabled remotely or through a specific combination of events (3.4.3)~\cite{koushanfar2011trusting}. Back-doors can also enable remote-access hence be detected in network traffic. The data obtained through the sensor should also be verified through data provenance (3.4.1), a malicious user could spoof sensors data, making end-users and operators believe in an event which is not occurring~\cite{xu2012data}. Physical devices can also be prone to misconfiguration leading to remote access, and data leaks (3.4.2).
\subsection{Human}
The Human factor also plays a decisive role in network security, without appropriate training and  awareness~\cite{bellekens2016study}. Humans can also be seen as the weak link, hence, when designing intrusion detection systems it is also important to identify  attacks against the end-users and expert-users~\cite{legg2016visual}. As shown in Figure~\ref{Fig:TaxonomyIntrusion}  (3.5.1) phishing and spear phishing attacks. These attacks are designed to trick the user to provide credentials or to allow a malicious user to access data without its consent. Some of these attacks may be targeted to have a higher probability of success. 

\subsection{Attack Examples}
This paper, aims at providing a simple and concise taxonomy of network attacks, and to this end, Figure~\ref{Fig:Examples} provides an example or recent attacks classified through our malicious taxonomy. These attacks include the Mirai botnet (1.1.1) that infected numerous IoT devices to launch distributed denial of service attacks against numerous Internet provided and services~\cite{kolias2017ddos}. It also includes the Tiny Banker Trojan (4.1), which was active from 2012 to 2016 targeting financial institutions stealing data from the users~\cite{marteau2017sequence}. Whilst these attacks are reported in the media, they are often excluded from training datasets. 

\section{Conclusion}
\label{conclusion}
In this paper a taxonomy of network threats for intrusion detection systems is presented. The taxonomy is divided into three control stages in order to describe more complex attack processes. The aim of this work is to create a taxonomy with the ability to inform researchers developing both intrusion detection systems and training datasets in order to increase the detection accuracy and decrease the false positive rate. With the increasing number of connected systems and networks the taxonomy aims at facilitating the design of future defense mechanisms as well as robust systems. 

\bibliographystyle{ieeetr}
\bibliography{bibliography}

\end{document}